\documentclass[sigconf]{acmart}
\AtBeginDocument{%
  \providecommand\BibTeX{{%
    \normalfont B\kern-0.5em{\scshape i\kern-0.25em b}\kern-0.8em\TeX}}}

\usepackage{multirow}
\usepackage{colortbl}
\newcommand{\method}{SemanticGNN~}
\newcommand{\company}{\textsc{Netflix}}

\begin{document}

\title{Synergistic Signals: Exploiting Co-Engagement and Semantic Links via Graph Neural Networks }

\author{Zijie Huang}
\affiliation{%
  \institution{University of California, Los Angeles}
  \country{}
  }
\author{Baolin Li}
\affiliation{%
  \institution{Northeastern University}
  \country{}
}
\author{Hafez Asgharzadeh}
\affiliation{%
  \institution{Netflix}
  \country{}
}
\author{Anne Cocos}
\affiliation{%
  \institution{Netflix}
  \country{}
}
\author{Lingyi Liu}
\affiliation{%
  \institution{Netflix}
  \country{}
}
\author{Evan Cox}
\affiliation{%
  \institution{Netflix}
  \country{}
}
\author{Colby Wise}
\affiliation{%
  \institution{Netflix}
  \country{}
}
\author{Sudarshan Lamkhede}
\affiliation{%
  \institution{Netflix}
  \country{}
}

\begin{abstract}
Given a set of candidate entities (e.g. movie titles), the ability to identify similar entities is a core capability of many recommender systems. Most often this is achieved by collaborative filtering approaches, i.e. if users co-engage with a pair of entities frequently enough, their embeddings should be similar. However, relying on co-engagement data alone can result in lower-quality embeddings for new and unpopular entities. 
We study this problem in the context recommender systems at \company{}. We observe that there is abundant semantic information such as genre, content maturity level, themes, etc. that complements co-engagement signals and provides interpretability in similarity models.
To learn entity similarities from both data sources holistically, we propose a novel graph-based approach called SemanticGNN.
SemanticGNN models entities, semantic concepts, collaborative edges, and semantic edges within a large-scale knowledge graph and conducts representation learning over it. Our key technical contributions are twofold: (1) we develop a novel relation-aware attention graph neural network (GNN) to handle the imbalanced distribution of relation types in our graph; (2) to handle web-scale graph data that has millions of nodes and billions of edges, we develop a novel distributed graph training paradigm. The proposed model is successfully deployed within \company{} and empirical experiments indicate it yields up to 35\% improvement in performance on similarity judgment tasks.
\end{abstract}

\keywords{Representation Learning, Graph Neural Networks, Knowledge Graphs}

\maketitle

\section{Introduction}

Recommender systems at \company{}  rely on entity (e.g. video title, game, actor) embeddings that accurately reflect the similarity between entity pairs. In many cases these representations are learned from user co-engagement data~\cite{liu2023www,pan2014sigir,Musa2020ICCTA,khojamli2021survey}: if two entities are frequently co-engaged by users, they should have similar embeddings. A well-known drawback of leveraging user collaborative signals in isolation to train entity embeddings is that these data suffer popularity biases and have low coverage for new and unpopular items~\cite{zhao2023www,lam2008icuimc}. At the same time, there may exist rich semantic information about items (such as genre, content maturity level, intellectual property, and storyline). This semantic data can provide an important signal in learning entity representation and also provides fine-grained interpretability for similarity scores.

\begin{figure}[t]
    \centering
    \vspace{3mm}
    \includegraphics[width=1.\linewidth]{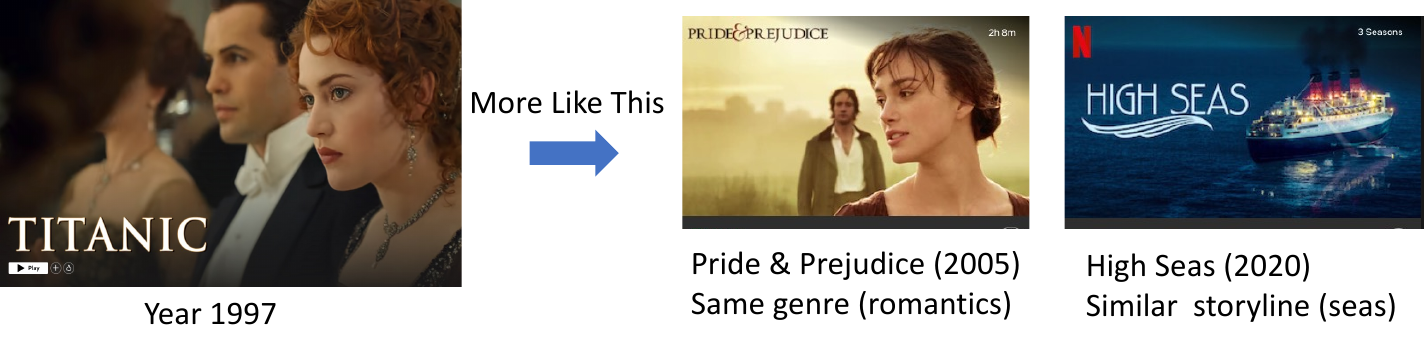}
    \vspace{-5mm}
    \caption{Similar entity generation in \company{}.}
    \label{fig:similar_titles}
\end{figure}

\begin{figure}[t]
    \centering
    \includegraphics[width=1.\linewidth]{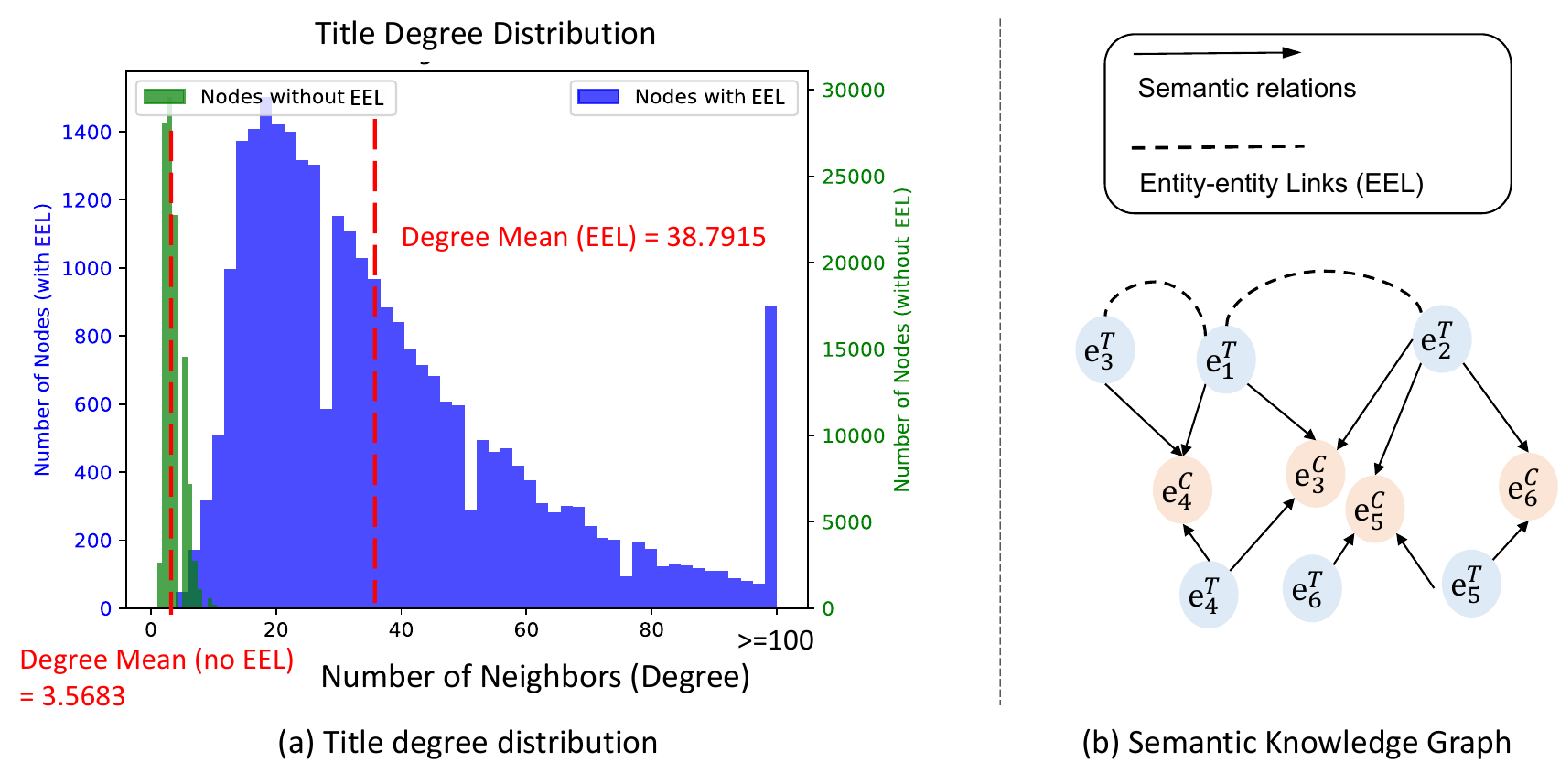}
    \caption{(a) Title degree distribution with/without entity-entity engagement links (EEL). The majority of titles are without EEL. (b) Constructed semantic knowledge graph for learning title representations.}
    \label{fig:motivation}
\end{figure}

Figure~\ref{fig:similar_titles} provides a concrete example. Based on co-engagement signals alone, a title similar to “Titanic” which was released in 1997 is “Pride \& Prejudice ”, which is also a popular romantic movie released in 2005. However, another more recent similar title is “High Seas” released in 2020. Compared with “Pride \& Prejudice”, it has a weaker co-engagement signal with "Titanic" but is semantically similar in the sense that both stories take place on the sea. Therefore, semantic information yields a more diverse set of similar titles and possibly helps us to understand \textit{why} two movies are similar, e.g. due to genre or storyline.

Motivated by these observations, we develop a novel method to learn entity embeddings from both co-engagement and content-based semantic information. Specifically, we train a Graph Neural Network (GNN)~\cite{GCN,GAT} over the knowledge graph (KG)~\cite{freebase,DBPedia} depicted in Figure~\ref{fig:motivation}(b). This enables us to capture both co-engagement and entity-specific semantic information in the GNN embeddings by modeling entities, semantic concepts (e.g. genres, maturity levels, storylines), and their relationships jointly.

GNNs~\cite{GCN,LGODE,hgt} have gained significant attention in recent years due to their superior ability to perform representation learning on graph-structured data~\cite{GAT,SSAGA,wu2022graph}. They have been successfully employed to perform industrial tasks such as building recommender systems~\cite{huang2021kdd,gao2023survey}, modeling user interest~\cite{qiu2022tkdd,shen2023temporal,DyDiffVAE}, etc. However, there are three unique challenges in training on this particular KG within \company{}, which existing GNN architecture designs are not fully aware of. 
\begin{enumerate}
    \item Relation type class imbalance: most entity nodes have semantic relations (i.e. links to semantic concept nodes), but entity-entity co-engagement links (EEL) between entity pairs are sparse as illustrated in Figure~\ref{fig:motivation}(a). Moreover, the number of semantic concepts associated with each entity also varies, resulting in imbalanced semantic edge distribution as well.
    \item Lack of informative features for semantic concept nodes: Unlike entities that are usually associated with abundant textual or visual features, semantic concepts are represented as short phrases that only contain limited information.
    \item Graph scalability issue: To handle web-scale graph data, we must design an efficient distributed training framework to train over the entire graph with millions of nodes and billions of edges on GPUs.
\end{enumerate}

To this end, we propose SemanticGNN, a large-scale end-to-end training pipeline for learning entity embeddings through a relation-aware GNN.  We design a two-step training strategy, where we first run pre-training via a KG completion task to generate contextualized representations of semantic nodes (e.g., genre), and second train a novel GNN model via link prediction loss utilizing the more sparse supervision signals from co-engagement relationships between entities. Specifically, the KG pre-training stage addresses challenge (ii) by forming representative embeddings for semantic concept nodes, in contrast to directly using concepts' short phrases as textual features. 

For challenge (i), we design a novel relation-aware GNN that is able to distinguish the influence of different neighbors of a node through attention weights. The attention weights are aware of different relation types such as \texttt{has\_genre} and \texttt{has\_maturity\_level}. In this way, for a newly released title that lacks any co-engagement data, we are able to distinguish the influence of different semantic types and thus learn an informative embedding. Distinguishing relation types also helps us to better represent popular titles: for a popular title that has abundant co-engagement links, thanks to the learnable prior weights of different relation types, the model is able to automatically adjust the influence received from co-engagement and from semantic edges, thus preventing noisy co-engagement data from dominating its representation. 

To address the scalability concerns in challenge (iii), we design a Heterogeneity-Aware and Semantic-Preserving (HASP) subgraph generation scheme to train our model across multiple GPUs -- such a framework has been proven efficient and feasible in an industry production environment. 
Our proposed method, SemanticGNN, is able to generate general-purpose entity representations that outperform baseline GNN-based entity embeddings in recognizing semantic similarity (35\% improvement) and co-engagement (21\% improvement). We also show that our method is especially powerful compared with baselines in the inductive setting and for learning new title representations. The model has also been successfully deployed within \company{}.

\section{Problem Definition}

A semantic knowledge graph describes entities and their associated semantic concepts such as genre. Formally, we represent a semantic knowledge graph as $\mathcal{G}(\mathcal{V}_t,\mathcal{V}_c,\mathcal{E}_{tc},\mathcal{E}_{tt})$, where $\mathcal{V}_t,\mathcal{V}_c$ are the sets of entity nodes and concepts nodes (e.g. \texttt{core\_genre}) respectively. The number of entity nodes is much larger than that of the concept nodes, i.e. $|\mathcal{V}_t|>>|\mathcal{V}_c|$. There are two relation sets: (1) $\mathcal{E}_{tc}$ are the directed entity-concept edges where each edge $e_{tc}$ points from an entity node $v_t$ to a concept node $v_c$. We denote $(v_t,e_{tc},v_c)$ as a semantic triple such as \texttt{(Titanic, has\_genre, romantic)}. We use $\mathcal{T}=\left\{(v_t,e_{tc},v_c)\right\}$ to denote the set of factual semantic triples. (2) $\mathcal{E}_{tt}$ are the undirected entity-entity links (EEL) obtained from user co-engagement data where if two titles are frequently co-engaged by users, an EEL would be created to denote their similarity. As a consequence of using user co-play data, such EELs are usually sparse and only cover a small portion of titles (biased toward popular titles) as shown in Fig~\ref{fig:motivation}(a). 

Given a semantic knowledge graph $\mathcal{G}(\mathcal{V}_t,\mathcal{V}_c,\mathcal{E}_{tc},\mathcal{E}_{tt})$, we would like to learn a model that effectively embeds entities to contextualized latent representations that accurately reflect their similarities. We evaluate the quality of the learned embeddings based on different entity pair similarity measurements, described in Section~\ref{sec:exp}.

\section{Related Work}
\subsection{Graph Neural Networks (GNNs)}
GNNs constitute a category of neural networks designed to directly process graph-structured data such as social networks~\cite{DyDiffVAE}, multi-agent dynamical systems~\cite{CGODE}, etc. They are applicable to diverse downstream tasks such as node classification~\cite{GCN,GAT}, graph clustering and matching~\cite{simgnn}, etc. Although diverse architectures exist, a typical update procedure for a single GNN layer involves two primary operations: (1) Information Extraction from Neighbors. In this step, information is gathered from each neighboring node. For instance, GAT~\cite{GAT} computes attention scores between node pairs based on sender and target node representations. These scores are then multiplied by the sender node's representation to yield the extracted information. GCN~\cite{GCN} employs the normalized Laplacian as the attention weight, treating it as an approximation of spectral domain convolution for graph signal processing. (2) Information Aggregation from Neighbors. Basic aggregation methods, such as mean, sum, and max, are commonly used. Additionally, more advanced pooling and normalization functions have been proposed.
Suppose $\boldsymbol{h}^l_t$ represents the node embedding for the target node $t$ during the ($l$)-th GNN layer. In this case, the update procedure from the ($l$-1)-th layer to the ($l$)-th layer can be succinctly described as follows:
\begin{equation}
\boldsymbol{h}^l_t \leftarrow \underset{\forall s \in N(t)}{\textbf { Aggregate }}\left(\textbf{Extract}\left(\boldsymbol{h}^{l-1}_s ; \boldsymbol{h}^{l-1}_t, e_{s\rightarrow t }\right)\right),
\end{equation}
where $N(t)$ denotes the neighbor (source) node sets of the target node $t$ and $e_{s\rightarrow t }$ denotes the edge type between pairs of the target node and source node.
Finally, through the accumulation of multiple layers (e.g. $L$ layers), each node's output representation, denoted as $\boldsymbol{h}^L_t$, incorporates messages from an expanded neighborhood. This results in a broader reception field, fostering high-order interactions among nodes and resulting in a  more contextualized node representation.

\subsection{Knowledge Graph Embeddings (KGE)}
Knowledge graph embeddings~\cite{transE,rotatE,ConvE,concept2box}  seek to acquire latent, low-dimensional representations for entities and relations,  which can be utilized to deduce hidden relational facts (triples). They measure triple plausibility based on varying score functions. For instance, translation-based models like TransE~\cite{transE} interpret relations as simple translations from a head entity to a tail entity. Extending this concept, TransH~\cite{TransH} projects entities into relation-specific hyperplanes, accommodating distinct roles of an entity in various relations. RotatE~\cite{rotatE}, in contrast, characterizes relations as rotations in the complex embedding space, capturing semantic properties like compositional relations. Later on, ConvE~\cite{ConvE} introduces a learnable neural network in computing the score function, and some recent studies explore leveraging textual input of entities and relations to enhance triple prediction tasks, integrating efficient language models like KG-Bert~\cite{KGbert}.

\subsection{Scaling up GNN Training}
\label{sec:related_scaling}

Training GNNs on large-scale graphs is a major bottleneck of today's GNN training when trying to take advantage of GPU acceleration~\cite{wang2021gnnadvisor,gandhi2021p3,zheng2022bytegnn}, as large graphs come with large memory and computation requirements, and requires system solutions to make these GNNs trainable on GPU memory and distributed across multiple GPUs~\cite{zhang2023survey}. Various system solutions tackle this problem by distributing the full graph across multiple GPUs and using inter-GPU communication to pass messages between nodes across different GPUs: Neugraph~\cite{ma2019neugraph} and Roc~\cite{jia2020improving} introduce distributed full-graph training by partitioning the graph across multiple GPUs and optimizing the communication and CPU/GPU memory management. Several following works have then focused on approximation and quantization to further improve distributed training speed~\cite{fey2021gnnautoscale,yu2022graphfm,wan2023adaptive}. In contrast to the full-graph approach, Pagraph~\cite{lin2020pagraph} and DistDGLv2~\cite{zheng2022distributed} use random sampling from the full graph to train on smaller graphs with optimized node caching and data movement. SALIENT~\cite{kaler2022accelerating} and BGL~\cite{liu2023bgl} have further optimized the dataloader and I/O of the sampling-based training process. However, these approaches are limited in the training of our SemanticGNN due to resource constraints and graph heterogeneity (Sec.~\ref{sec:deployment}).
\section{Method}

\begin{figure*}[t]
    \centering
    \includegraphics[width=0.8\linewidth]{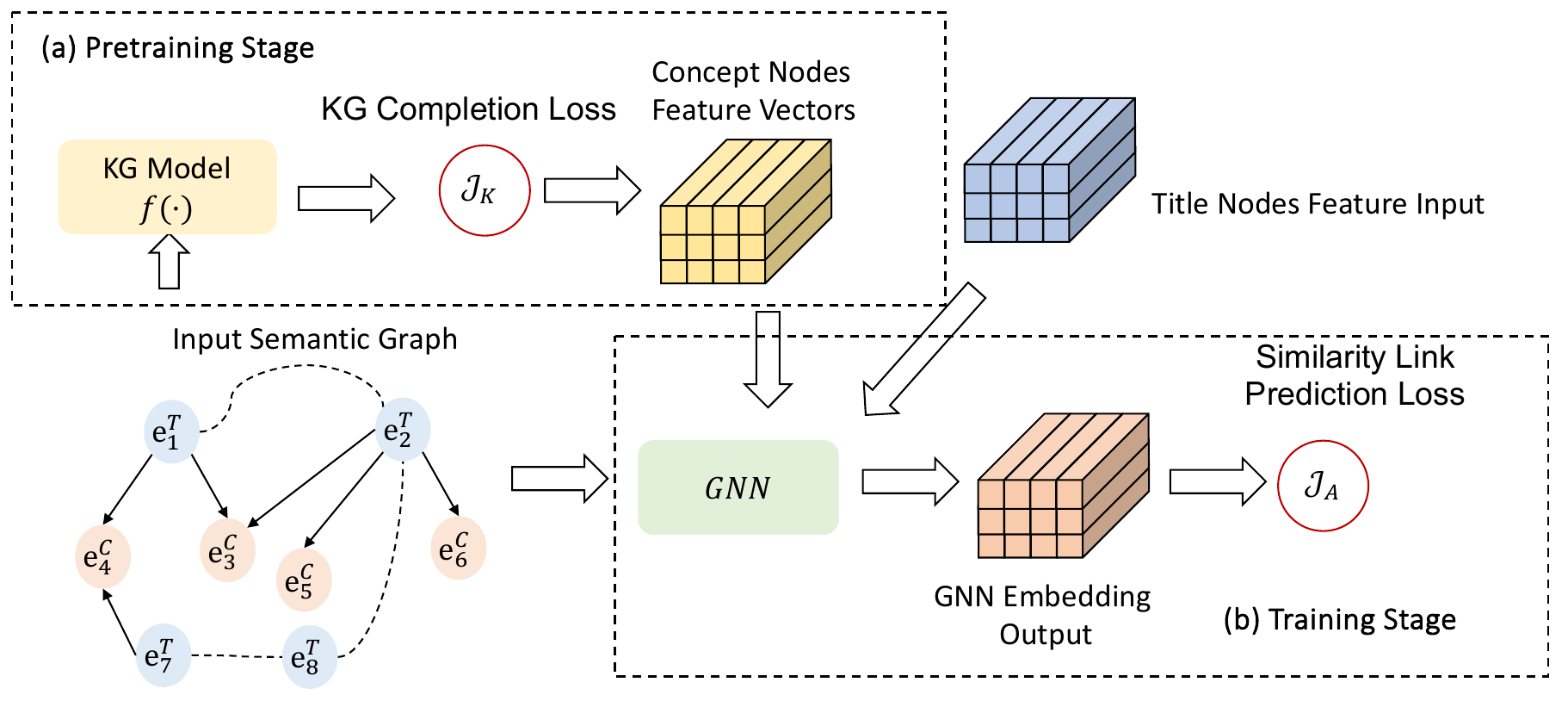}
    \caption{Overall framework of SemanticGNN: (a) KG pretraining for getting embeddings for concept nodes. (b) Training stage based on relation-aware GNN and link prediction loss on EEL.}
    \label{fig:method}
\end{figure*}

Figure~\ref{fig:method} shows the overall framework of our proposed SemanticGNN. It follows a two-step training pipeline: 1.) We first pretrain the semantic KG via KG completion loss using TransE in order to generate the embeddings for concept nodes. 2.) We then train a relation-aware GNN over the KG using link prediction loss over EEL to get entity embeddings that reflect entity similarity. Now we introduce each component in detail.

\subsection{KG Pretraining}

Our goal in KG pretraining is to produce high-quality features for semantic concept nodes, as concept nodes are usually associated with short phrases, which may not be informative enough to serve as input features. We choose TransE~\cite{transE} as our backbone and conduct KG pretrating via the standard KG completion task~\cite{concept2box}. It can be replaced with any off-the-shelf KG embedding method based on different downstream applications and KG structures. Specifically, let $\boldsymbol{e}_t, \boldsymbol{e}_c$ be the learnable embeddings of entity $t,c$ respectively, we train entity embeddings via the hinge loss over semantic triples $\mathcal{T}=\left\{(v_t,e_{tc},v_c)\right\}$ defined as: 
\begin{equation}
\mathcal{J}_K=\sum_{\mathcal{T}}\left[f\left(\boldsymbol{e}_t^{\prime}, \boldsymbol{r}_{tc}, \boldsymbol{e}_c^{\prime}\right)-f\left(\boldsymbol{e}_t, \boldsymbol{r}_{tc}, \boldsymbol{e}_c\right)+\gamma\right]_{+},
\end{equation}
where $\gamma>0$ is a positive margin, $f$ is the KGE model, and $\boldsymbol{r}_{tc}$ is the embedding for the relation $e_{tc}$. $(\boldsymbol{e}_t^{\prime},\boldsymbol{r}_{tc},\boldsymbol{e}_c^{\prime})$ is a negative sampled triple
obtained by replacing either the head or tail entity of
the true triple $(\boldsymbol{e}_t, \boldsymbol{r}_{tc},\boldsymbol{e}_c)$ from the whole entity pool.

\subsection{Relation-Aware GNN}
To handle the imbalanced relation distribution in the constructed KG, we propose an attention-based relation-aware GNN to learn contextualized embeddings for entities following a multi-layer message passing architecture.

In the $l$-th layer of GNN, the first step involves calculating the relation-aware message transmitted by the entity $v_t$ in a relational fact $(v_t,e_{tc},v_c)$ using the following procedure:

\begin{align*}
    \boldsymbol{h}_{c(r)}^l = {\rm Msg}\left(\boldsymbol{h}_{c}^l, \boldsymbol{r}\right)  := \boldsymbol{W}^l_v {\rm Concat}(\boldsymbol{h}_i^l, \boldsymbol{r}),
\end{align*}
where $\boldsymbol{h}_{c(r)}^l$ is the latent representation of $v_c$ under the relation type $r$ at the $l$-th layer, ${\rm Concat}(\cdot,\cdot)$ is the vector concatenation function, $\boldsymbol{r}$ is the relation embedding and $\boldsymbol{W}^l_v$ is a linear transformation matrix.
Then, we propose a relation-aware scaled dot product attention mechanism to characterize the importance of each entity's neighbor to itself,
which is computed as follows: 

\begin{align}
    &{\rm Att}\left(\boldsymbol{h}_{c(r)}^l,\boldsymbol{h}_t^l\right) = \frac{\exp ( \alpha_{ct}^{r} ) }{\sum_{v_{c^{\prime}}\in \mathcal{N}(v_t) } \exp \left(\alpha_{c^{\prime}t}^{r}\right)} \notag
    \\
    &\alpha_{ct}^{r} = \left(\boldsymbol{W}_k^l \boldsymbol{h}_{c(r)}^l \right)^T \cdot \left(\boldsymbol{W}_q^l \boldsymbol{h}_t^l\right)\cdot \frac{1}{\sqrt{d}}\cdot \beta_r, 
    \label{eq:attention}
\end{align}
where $d$ is the dimension of the entity embeddings, $\boldsymbol{W}_k^l, \boldsymbol{W}_q^l$ are two transformation matrices, and $\beta_r$ is a learnable relation factor for each relation type $r$. 
Diverging from conventional attention mechanisms~\cite{GAT,simgnn}, we incorporate $\beta_r$ to represent the overall significance of each 
relation type $r$. This is crucial, as not all relations equally contribute to the targeted entity depending on the overall KG structure.

We then update the hidden representation of entities by aggregating the message from their neighborhoods based on the attention score:

{\small
\begin{align*}
    \boldsymbol{h}_t^{l+1} =  \boldsymbol{h}_t^{l} + \sigma\left(\sum_{v_c\in \mathcal{N}(v_t)}{\rm Att}\left(\boldsymbol{h}_{c(r)}^l,\boldsymbol{h}_t^l\right)\cdot \boldsymbol{h}_{c(r)}^l\right),
\end{align*}}%
where $\sigma(\cdot)$ is a non-linear activation function, and the residual connection is used to improve the stability of GNN~\cite{resnet}.
Finally, we stack $L$ layers to aggregate information from multi-hop neighbors and obtain the final embedding for each entity $i$ as $\boldsymbol{h}_i = \boldsymbol{h}_i^{L}$. 

\subsection{Overall Training}

Given the contextualized entity embeddings, we train the model using standard link prediction loss defined over entity-entity links:

{
\small
\begin{align}
    &\mathcal{J}_{A}\hspace{-0.05in}= -\frac{1}{N} 
\sum_{\substack{e_{ij}\in\mathcal{E}_{tt}\\e_{ij'}\notin\mathcal{E}_{tt}}} \log \left(p\left(e_{ij}\right)\right)+ \log \left(1-p\left(e_{ij'}\right)\right),\\
&\text{where}~\quad p(e_{ij}) = \text{Sigmoid}(\boldsymbol{h}_i^T\cdot \boldsymbol{h}_j).
    \label{eq:LP_loss}
\end{align}
}%

\section{System Deployment}

Next, we describe a practical distributed GNN training framework that addresses the challenge of scalability and practicality in training GNNs for deployment. Traditional GNN training methodologies often hit bottlenecks such as memory limitations and computational inefficiencies, particularly when the task involves large-scale graphs with complex relations and high-dimensional feature spaces. While CPU-based training solutions exist, they prove to be prohibitively slow for graphs of this scale and complexity (taking days to train). On the other hand, training these large graphs solely on GPUs is also infeasible due to limitations in GPU DRAM. Our framework ameliorates these challenges by training on a set of sub-graphs with preserved semantic information and distributing the computational workload across multiple GPUs in a cluster. This distributed approach allows for parallel processing and significantly reduces per-GPU memory consumption, thereby accelerating the training process. It does so without sacrificing the model's ability to learn entity embeddings using semantic information, effectively bridging the gap between computational efficiency and model performance.

\begin{figure*}[ht]
    \centering
    \includegraphics[width=0.96\linewidth]{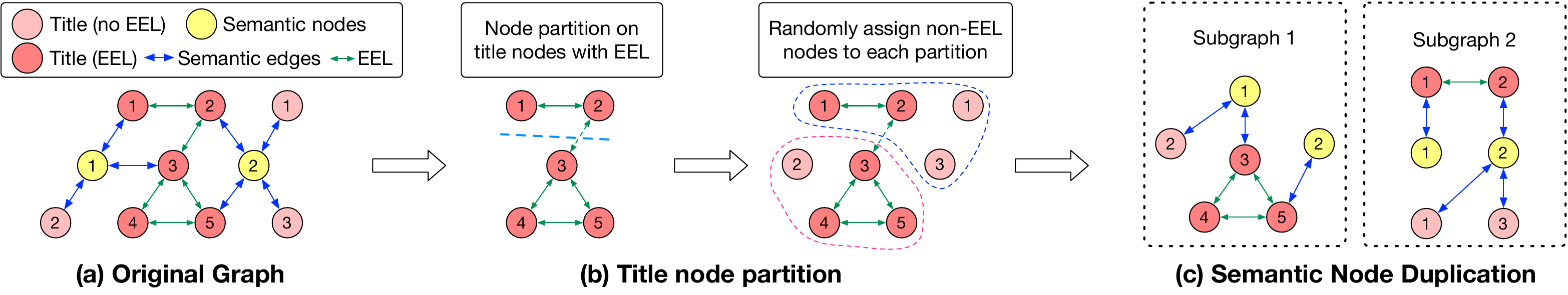}
    \caption{Demonstration of the subgraph generation process on a two-subgraph example.}
    \label{fig:method_system}
\end{figure*}

\subsection{Training Challenges}
\label{sec:deployment}

Deploying the GNN training workflow on a multi-GPU system presents a formidable challenge, primarily stemming from the stringent limitations of GPU device memory and inter-GPU communication. Our input graph boasts a substantial number of entity nodes, each accompanied by high-dimensional node features. Moreover, Equation ~\ref{eq:attention} highlights another storage-intensive requirement, as it necessitates the retention of computed attention coefficients for every pair of nodes connected by an edge. In our graph, these edges represent user co-play history, resulting in a significant volume of entity-entity connections. To put this into perspective, even on a powerful NVIDIA A100 Tensor Core GPU with 40GB of memory, we can only feasibly train a semantic knowledge graph of a certain size. Yet, our target graph encompasses 7$\times$ more entity nodes and 7$\times$ more edges than the largest graph we can train on a single A100. Furthermore, our knowledge graph is poised for continual expansion, demanding a training framework that remains future-proof in the face of these escalating memory demands.

Distributed training of our knowledge graph over multiple GPUs presents a unique set of challenges that are currently unaddressed by existing off-the-shelf solutions. As discussed in Sec.~\ref{sec:related_scaling}, previous multi-GPU GNN training techniques can be categorized into two approaches: (i) the full-graph training approach and (ii) the graph sampling approach. However, neither of these approaches can be well applied to our use case as we discuss in the following.

While full-graph training preserves cross-GPU edge connections through inter-GPU communication during training, it also poses limitations. Specifically, the number of GPUs required scales with the graph size, making it a less flexible option for our dynamic workloads at \company{}. In our setting, we aim for a system that can train on an arbitrary number of GPUs, regardless of whether there are, for example, at least 16 GPUs available in the system. 

The graph sampling approach hosts the graph in CPU memory and samples random nodes to create subgraphs, which are then loaded into the GPU for training. However, this approach is ill-suited for our use-case due to several reasons: First, our semantic knowledge graph is highly heterogeneous, with each node type having different feature sizes and degrees. Existing solutions either focus on homogeneous graphs or convert heterogeneous graphs into homogeneous ones \cite{zheng2022distributed}. Secondly, the graph sampling and streaming from the CPU into the GPUs can become a performance bottleneck during training. Finally, Our application features high-degree nodes, such as one semantic node connected to tens of thousands of entity nodes. The sampling strategy would thus either capture an excessively large subgraph or require us to discard many neighbors to fit into the GPU memory. Given these challenges, there is a clear need for a more adaptable and scalable multi-GPU GNN training solution for heterogeneous graphs.

\subsection{HASP Multi-GPU Training Framework}
\label{sec:distributed_framework}

We propose a novel and practical framework, termed Heterogeneity-Aware and Semantic-Preserving (HASP), for generating subgraphs from a comprehensive semantic knowledge graph. Unlike traditional methods such as graph sampling, where nodes are randomly selected, or graph partitioning, which divides the entire graph into several subgraphs, HASP is designed to capture the graph heterogeneity and preserve the semantics within the generated subgraphs. The advantages of the HASP approach are manifold. Firstly, each HASP-generated subgraph is optimized to fit within the memory constraints of modern GPUs, ensuring efficient training without memory overflow issues. Secondly, it offers both data parallelism and flexibility: enabling simultaneous GNN training on these subgraphs across multiple GPUs, while also accommodating scenarios where a single GPU can sequentially process the HASP subgraphs using a dataloader. Next, we explain the design details of the HASP generation framework, illustrated in Fig.~\ref{fig:method_system}.\vspace{2mm}

\noindent\textbf{Entity node partitioning.} In the HASP framework, the entity nodes within each subgraph are distinct, ensuring no overlap across subgraphs. To achieve this, a specific subset of entity nodes is allocated to each HASP subgraph. Notably, relationships between entity nodes are represented by entity-entity links (EELs), indicative of relationships like user co-play. A challenge arises when dividing entity nodes: if two nodes connected by an EEL are apportioned to separate HASP subgraphs, the EEL is effectively bypassed during GNN training. This diminishes the utility of such relationships in the learning process. Furthermore, to avoid computational disparities and ensure efficient distributed training, it is imperative that each HASP subgraph contains a roughly equal count of entity nodes. Uneven distribution might lead to GPU workload imbalances and resource underutilization. Therefore, our entity node partitioning strategy strives to achieve two primary objectives: (i) minimizing the elimination of EELs across HASP subgraphs and (ii) ensuring an equitable distribution of entity nodes across subgraphs.
\begin{table*}[htbp]

\caption{Evaluation results of baselines and our proposed method \method over two ground-truth entity similarities: HC-Sim (human-curated semantic sims) and Co-Sim (observed co-engagements). We report the relative improvement ratio compared with \company{}-baseline.}

\label{tab:overall_results}
\centering
\scalebox{1}{
\begin{tabular}{lcccccc}
\hline
\multicolumn{1}{c|}{Models\textbackslash{}}                                                  & \multicolumn{3}{c|}{Co-Sim}            & \multicolumn{3}{c}{HC-Sim}      \\
\multicolumn{1}{c|}{Metrics}                                                                 & MAP@10 & MAP@50 & \multicolumn{1}{c|}{MAP@100} & MAP@10                        & MAP@50                         & \multicolumn{1}{c}{MAP@100}            \\ \hline
\multicolumn{7}{c}{\cellcolor{gray!25}\textbf{Baseline Methods}}    \\ \hline      
\multicolumn{1}{l|}{GCN}                                                                     & -55.34\% & -60.98\% & \multicolumn{1}{c|}{-62.26\%}  & -49.32\%                       & -54.26\%                         & \multicolumn{1}{c}{-55.53\%}      \\
\multicolumn{1}{l|}{GAT}                                                                     & -15.96\%     & -12.61\%       & \multicolumn{1}{c|}{-13.23\%}        &    -35.15\%                      &            -33.07\%                   & \multicolumn{1}{c}{-32.93\%}                                            \\
\multicolumn{1}{l|}{GraphSAGE}                                                               &     -12.01\%   &   -9.19\%     & \multicolumn{1}{c|}{-10.31\%}        &    -30.25\%                           &   -25.32\%                            & \multicolumn{1}{c}{-27.64\%}                           \\
\multicolumn{1}{l|}{\begin{tabular}[c]{@{}l@{}}\company{}-baseline\end{tabular}} & - & - & \multicolumn{1}{c|}{-}  & -                       & - & \multicolumn{1}{c}{-}                             \\ \hline
\multicolumn{7}{c}{\cellcolor{red!25}\textbf{Proposed Method}}\\                    
\hline 
\multicolumn{1}{l|}{SemanticGNN\_no\_KG}                                                       & 18.59\% & 13.05\% & \multicolumn{1}{c|}{12.54\%}  & 30.79\% &  27.13\%  & \multicolumn{1}{c}{ \textbf{26.20\%}}  \\
\multicolumn{1}{l|}{SemanticGNN}                                                             & \textbf{21.67\%} &  \textbf{17.06\%} & \multicolumn{1}{c|}{ \textbf{18.25\%}}  & \textbf{ 35.42\%}                        &  \textbf{29.46\%} & \multicolumn{1}{c}{25.96\%}                        \\ \hline
\end{tabular}}
\end{table*}

Our partitioning methodology begins by categorizing entity nodes of the original graph into two distinct groups: nodes that possess EELs and nodes that lack any EELs, as illustrated in Fig.~\ref{fig:method_system} (a). For the former group, we engage in a minimum cut graph partitioning process on the EEL subgraph, which is composed of all EELs and their interconnected nodes. The METIS library\cite{karypis1997metis} facilitates this operation, delivering $N$ approximately uniform partitions of the EEL subgraph with minimized EEL eliminations. Here, $N$ stands as a representation of our predetermined number of target partitions, generally set to be a multiple of available GPUs and optimized in line with the GPU memory capacity to ensure each HASP subgraph can comfortably reside within GPU memory. Following this partitioning, we then integrate nodes without EELs into these $N$ partitions. Allocation begins from the most sparsely populated partition, ensuring a balanced distribution of entity nodes across all partitions, thus equalizing computational load. This method allows us to achieve the dual goals of minimal EEL elimination and equitably distributed entity nodes. This entity node partition process is demonstrated in Fig.~\ref{fig:method_system} (b). We deliberately avoided partitioning the entire entity node subgraph, including both the EEL and non-EEL nodes, in one go to prevent skewed EEL distributions, where some partitions could be densely populated with nodes interconnected by EELs while others might be bereft of them, potentially undermining the generality of our resulting HASP subgraphs.\vspace{2mm}

\noindent\textbf{Semantic node duplication. }  In our HASP generation process, each subgraph encompasses an exclusive set of entity nodes. Yet, to ensure the preservation of all semantic information, every subgraph retains a full set of semantic nodes, effectively duplicating them across all subgraphs. This design choice, illustrated in Fig.~\ref{fig:method_system} (c), is motivated by the vast difference in quantity between entity and semantic nodes. To elucidate, consider the graph's representation of genres: there might only be 20 distinct genre types, each represented by a single semantic node. In contrast, each genre could be associated with tens of thousands of \company{} shows (the entity nodes), highlighting the stark disparity in node counts. This approach ensures all the semantic information can be used in the message passing when training the GNN on each individual subgraph. \vspace{2mm}

\noindent\textbf{User-defined node sampling. } In some scenarios, the distinction between entity nodes and semantic nodes may blur. For instance, when integrating external show entities into the graph to enrich semantic information, these entities don't neatly fit into the traditional semantic node category. Given their potentially vast quantities, duplicating these external entities, akin to the way we handle traditional semantic nodes, becomes impractical. To address this, our HASP framework introduces a user-defined node sampling feature. This allows users to specify and randomly sample a set number of nodes from particular node types. This inclusion offers context within the generated subgraphs while ensuring we remain within memory constraints. 

\vspace{2mm}

In summary, the HASP framework adeptly generates subgraphs from a semantic knowledge graph. These subgraphs can be subsequently trained on an arbitrary number of GPUs or even sequentially on a single GPU. By duplicating semantic nodes across all subgraphs, HASP ensures the consistent preservation of semantics. Title nodes, on the other hand, are strategically partitioned across subgraphs. This ensures their distinctiveness while taking into account their intricate interconnections, represented as EELs, and simultaneously achieving a balanced distribution. For complex scenarios like integrating external show entities, HASP provides a user-defined node sampling, allowing customized node inclusion while staying within memory limits. Users can leverage the HASP framework for efficient GNN training as we have observed \textit{an average improvement of over 50$\times$} when trained on 4 A100 GPUs compared to a naive full-graph training on a large-memory CPU machine. For inference, given its one-time effort nature, node embeddings can be generated using a large-memory CPU machine, ensuring all edges in the knowledge graph are holistically preserved.

\section{Experiments}~\label{sec:exp}
\subsection{Experiment Setup}

\textbf{Datasets.} We constructed our semantic knowledge graph by collecting entities and their associated metadata from \company{}'s catalog. To empirically evaluate our model performance, we subsample a portion of it and in total, we have over 125k nodes, where nearly 99\% are entities and 1\% are semantic concepts. Semantic concepts describe basic information associated with each entity such as genre and in total we have roughly 10 different categories of such semantic concepts. 
For edges, we have both entity-entity edges (EEL) with limited amounts and abundant entity-semantic edges. The total number of edges is around $10\times$ compared to the total number of nodes, where around 90\% of them are entity-semantic edges. When developing the model at~\company{}, it handles graph size to around 1 million entity nodes following a similar structure. We only report the results on the sampled KG in this paper. In practice, the deployed model over the full KG shows great performance gain in varying downstream recommendation tasks, especially for new/unpopular titles.


\subsection{Implementation Details.}
\textbf{Training Details.} Our codebase is written in Python 3.7.12, we use Pytorch 1.12.0 as the training framework and use the Pytorch Geometric 2.2.0 library to construct the graph. We perform GNN training on our Ray~\cite{moritz2018ray}-based machine learning training platform using NVIDIA A100 Tensor Core GPUs, each with 40GB GPU memory. The CUDA version is 11.4. \newline

\noindent\textbf{Evaluation Protocol. }
We evaluate the quality of the learned entity embeddings through two offline evaluation benchmarks, which cover two flavors of entity-entity similarity measurements.
\begin{itemize}
    \item Co-engagement Similarities (Co-Sim): Observed pairs of entities in which both entities were engaged by the same user.
    \item Human-curated Similarities (HC-Sim): Dataset of similar entity pairs, annotated by human experts.
\end{itemize}
\vspace{2mm}

\noindent\textbf{Baselines. } We compared against GCN~\cite{GCN}, GAT (heto version~\footnote{The heterogeneous version implemented in the pytorch geometric library \href{https://pytorch-geometric.readthedocs.io/en/latest/notes/heterogeneous.html}{here}.})~\cite{GAT}, GraphSAGE (heto version)~\cite{hamilton2017inductive}, and \company{}-baseline model. 
\company{}-baseline model adopts GAT (heto version) + embedding similarity loss for training. The embedding similarity loss is designed as cosine similarity measurements between the learned embeddings to other pre-trained embeddings that also reflect title similarities from other data sources (with limited coverage of titles in the constructed KG), with negative sampling. It is in general more time-expensive compared with the link prediction loss in SemanticGNN.

\subsection{Main Results}

We first examine whether our SemanticGNN has superior performance compared with baselines as well as the design rationality of each component of SemanticGNN.
As illustrated in table~\ref{tab:overall_results}, our proposed SemanticGNN in general is able to outperform baselines across evaluation metrics. Our model variant without KG pretraining has degraded performance compared with SemanticGNN, showing the effectiveness of designing the KG pre-training stage.

\subsection{Results on different entity groups based on their degree}

\begin{figure}
    \centering
    \includegraphics[width=1\linewidth]{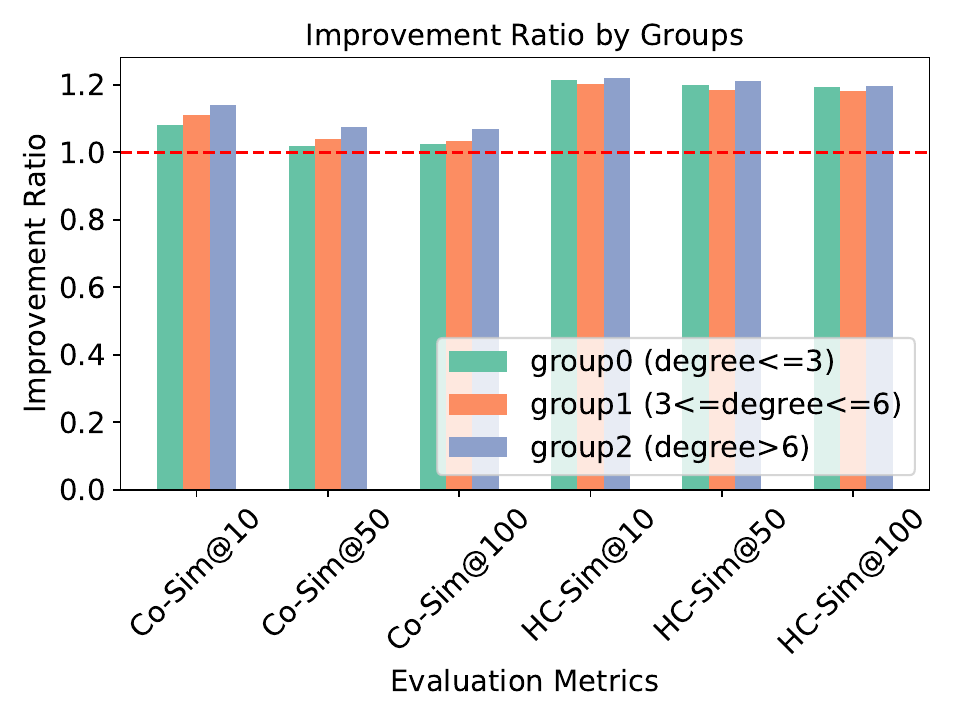}
    \caption{Model performance compared with \company{}-baseline over three different entity groups.}
    \label{fig:grouping_main}
\end{figure}

We next show more fine-grained evaluation results by grouping our entities into three groups based on their EEL degree within the graph, which is a rough proxy for popularity. We delineate \textit{group0} (degree$\leq 3$), \textit{group1} ($3<$degree$<=6$) and \textit{group2} (degree$>6$). The number of entities within each group are 60k, 45k, and 33k, which are relatively balanced. We would like to investigate how our model performs on popular entities and new entities in this experiment. We compare our method with \company{}-baseline which is the strongest baseline we have. Specifically, we compute the improvement ratio of SemanticGNN/\company{}-baseline across groups and evaluation metrics. The results against three groups are shown in Figure~\ref{fig:grouping_main}.

We can firstly observe that SemanticGNN is able to surpass \company{}-baseline across groups, showing its superior performance among varying entity popularities. Also, SemanticGNN is more helpful towards \textit{group0} (least popular) and \textit{group2} (most popular) against baselines. This may indicate that our way of incorporating semantic information is in general helpful for learning nodes that are solely based on semantic information (\textit{group0}) and nodes that have strong signals in user co-engagement data (\textit{group2}).

\subsection{Results of Inductive Setting}
We next would like to test our model performance in the inductive setting: if we only train over 80$\%$ of the total entities in the graph and evaluate over the leftover 20$\%$ entities in our graph, how would it perform? This would be a more realistic setting when we would like to use the model in production: when a new entity comes, we do not need to retrain the model, but instead just plug the added nodes and edges into the graph, and run the model for inference. We again report the results across the three groups. Our train/test split is conducted over the three groups and then union them as a whole.

\begin{figure}
    \centering
    \includegraphics[width=1\linewidth]{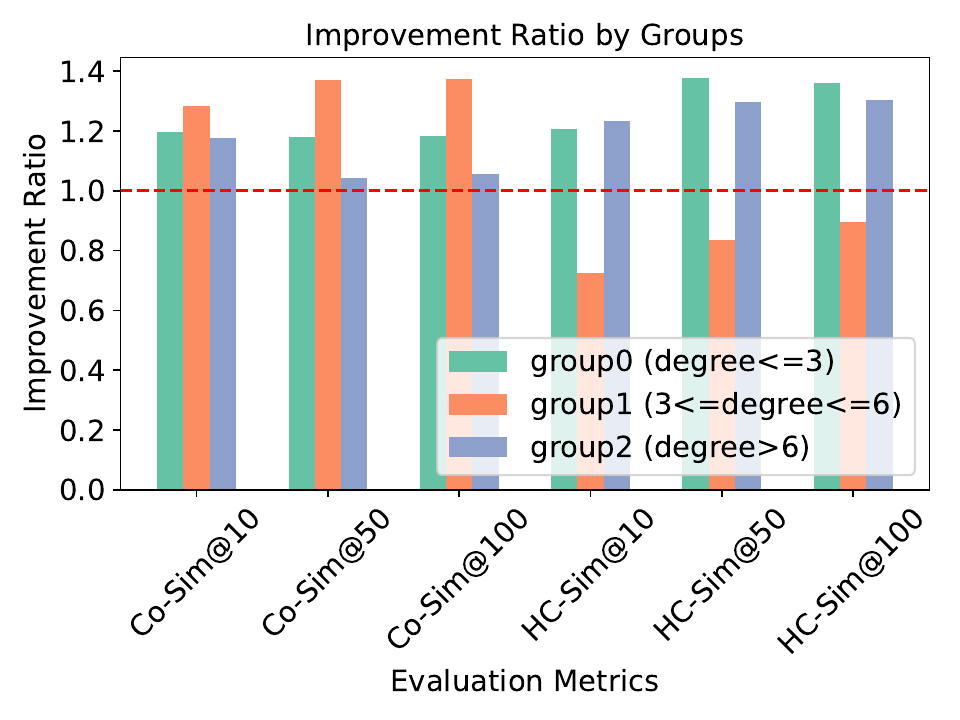}
    \caption{Model performance compared with \company{}-baseline over three different entity groups in the inductive setting.}
    \label{fig:grouping_edges}
\end{figure}

As shown in Figure~\ref{fig:grouping_edges}, we can observe SemanticGNN has better performance compared with \company{}-baseline in the inductive setting. It indicates that SemanticGNN is more generalizable and has better mechanisms for learning and aggregating different edge types. By computing the improvement ratio as SemanticGNN/\company{}-baseline shown below, we found in general our method is more beneficial to new entities (group0).

\subsection{Results of Adding New EEL edges during evaluation}
As mentioned previously, the majority of the entities (shown in~\ref{fig:motivation}(b)) are without EELs and are only associated with semantic information. However, as we identify that EEL in general serves as a stronger signal for learning entity similarities, we wonder whether we can utilize our learned embeddings to identify more EELs, and then utilize them to revise the graph structure so as to get better results.

\begin{figure}
    \centering
    \includegraphics[width=1\linewidth]{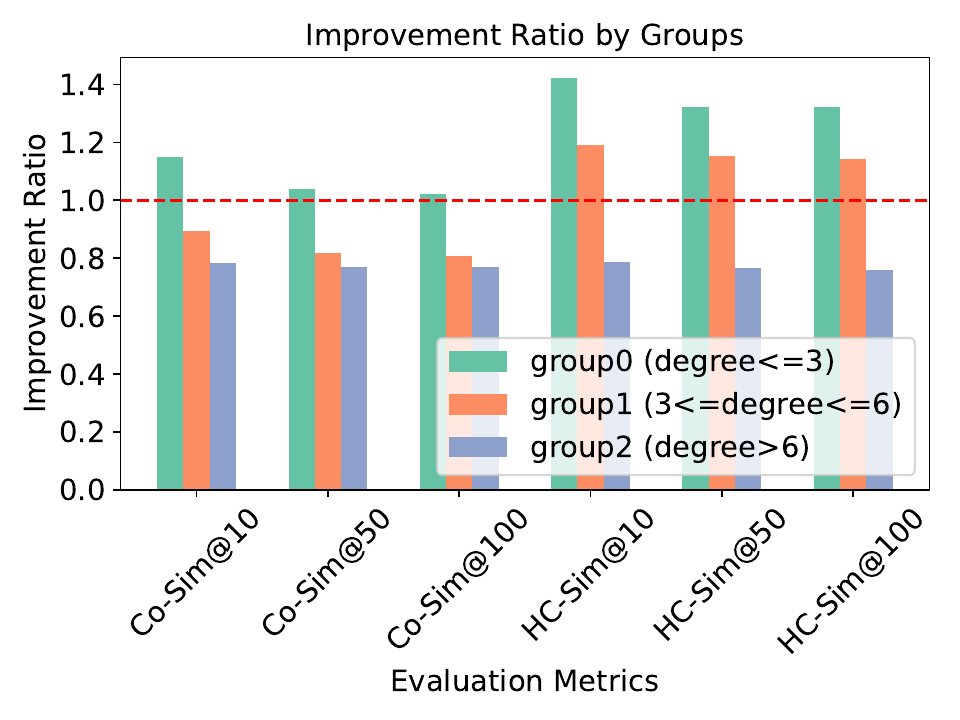}
    \caption{Model performance over three different entity groups when adding new edges.}
    \label{fig:grouping_newEEL}
\end{figure}

Specifically, we utilize our output entity embeddings from SemanticGNN to compute the top 3 nearest neighbors for each entity based on their L2 distance. We inject these new EELs into the graph and use the learned model to do inference on the revised graph (without retraining the model). The original graph has 482,031 EELs in total, after revising it, it has 845,732 EELs. We show our results by groups as in Figure~\ref{fig:grouping_newEEL}.

We can observe that adding new EEL edges improves the performance of group0 which is the set of least popular entities. However, it harms the performance slightly for group1 and group2. One likely interpretation is that the new EELs are most beneficial for new entities that previously only received information from semantic nodes. For popular entities, adding too many EELs may inject noise into their representations. One way to avoid this is during message passing, we keep new EELs directed, i.e. there would only be messages delivered from popular entities to new entities, without new entities to popular entities (as popular entities already have very dense EELs).

\subsection{Sensitivity to Distributed Training}

\begin{figure}
    \centering
    \includegraphics[width=1\linewidth]{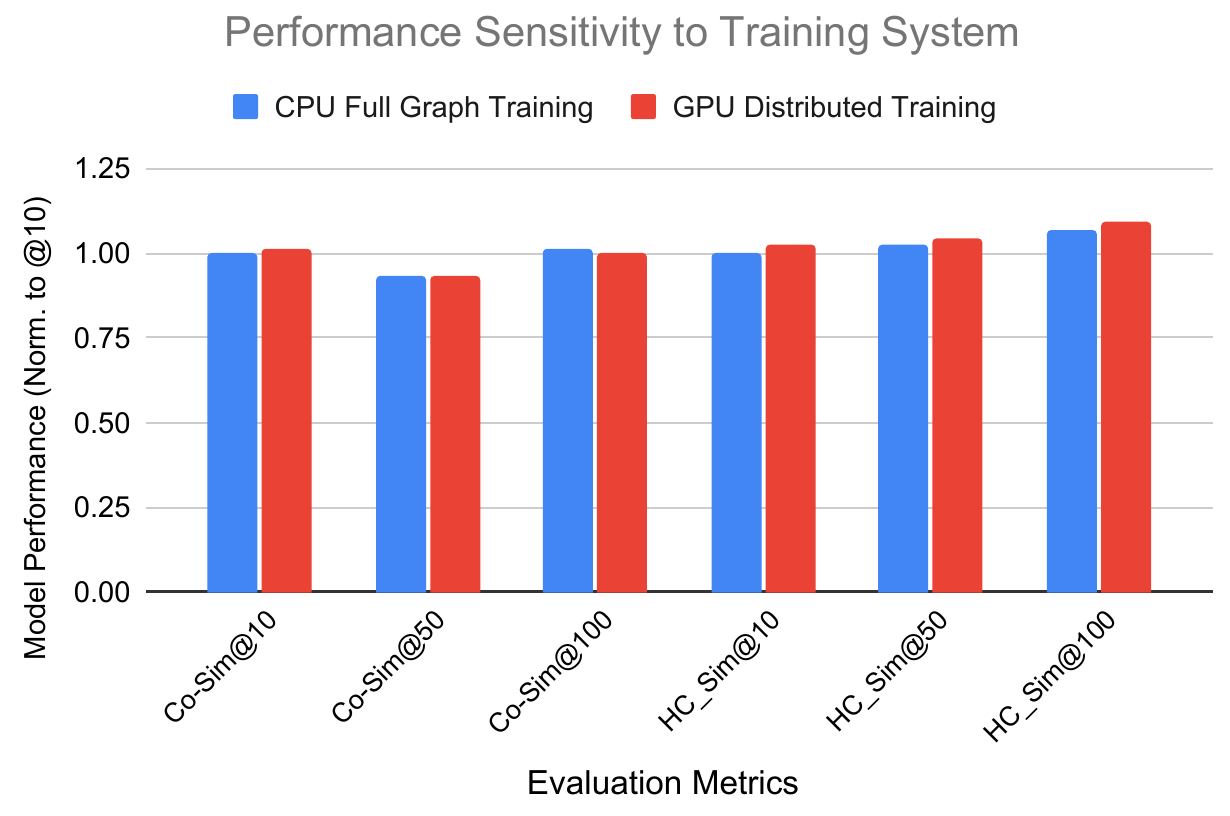}
    \caption{GNN performance sensitivity to distributed training over HASP-generated subgraphs.}
    \label{fig:eval_partition}
\end{figure}

Recall that in Sec.\ref{sec:deployment}, we introduced a distributed training framework that generates a series of subgraphs. Each subgraph is designed to fit within the GPU memory and is trained independently. A natural question arises: How does the performance of this subgraph training approach compare to full-graph training, which is restricted to CPUs due to memory constraints? To answer this question, we compare the models trained by our subgraph-based distributed training method against the CPU-only full-graph training. Note that in the CPU-only training, we keep the same graph and GNN architecture, but train the GNN on a memory-optimized CPU node for the same number of epochs, which takes more than a day. 
Fig.~\ref{fig:eval_partition} provides a reassuring answer as training the GNN using multiple GPUs results in similar model performance as training over the full graph on CPU. This success is attributed to our meticulous design strategies, including entity node partitioning and semantic node duplication, as detailed in Sec.~\ref{sec:distributed_framework}. This emphasizes the advantage of the distributed training framework, as not only does our approach offer significant speedup, but it also ensures that there's no compromise on model performance throughout the training pipeline.

\section{Conclusion and Future Work}
In this work, our goal is to effectively use co-engagement signals and entity metadata to improve cold-starting performance on the entity-entity similarity task. We proposed a novel solution SemanticGNN and showed its effectiveness on \company{}'s entity-entity similarity problem. This work presents effective solutions to the challenges presented by the semantic KG: first, the heterogeneity and imbalanced edge distributions among nodes are addressed via a relation-aware GNN architecture; second, we learn informative node features for semantic concept nodes via KG-pretraining; and third, we scale training over a graph with millions of nodes and billions of edges via our HASP distributed multi-GPU training framework. Our evaluation results suggest that:
\begin{itemize}
\item SemanticGNN is able to outperform baselines by a large margin while costing less computational time
\item SemanticGNN is especially useful for new entities and in the inductive setting compared with baselines
\item SemanticGNN has the potential to generate more EEL edges that can further enhance model performance.
\end{itemize}
To advance our understanding of semantic relationships, we are interested in investigating the hierarchical structuring of concepts in the future. For instance, broad categories like "person" may encompass more nuanced sub-concepts such as "singer". Developing models that capture these hierarchies could significantly enhance our semantic analysis capabilities.

\bibliographystyle{ACM-Reference-Format}
\bibliography{sample-base}

\appendix


\end{document}